\documentstyle[aps,prd,floats,psfig]{revtex}

\draft

\newcommand{\be}{\begin{equation}}
\newcommand{\ee}{\end{equation}}
\newcommand{\bea}{\begin{eqnarray}}
\newcommand{\eea}{\end{eqnarray}}
\newcommand{\bml}{\begin{mathletters}}
\newcommand{\eml}{\end{mathletters}}
\renewcommand{\vec}[1]{{\bf #1}}

\begin{document}

\wideabs{
\title{Fluids of Vortices And Dark Matter}
\author{Filipe Bonjour\cite{bonj} and P.S. Letelier\cite{psl}}
\address{Instituto de Matem\'atica, Estat\'{\i}stica e Computa\c c\~ao
  Cient\'{\i}fica (IMECC),\\ Universidade Estadual de Campinas, CP 6065,
  13081-970 Campinas SP, Brazil}
\setlength{\footnotesep}{0.5\footnotesep}
\maketitle

\begin{abstract}
  By considering full-field string solutions of the Abelian--Higgs model, we
  modify the model of a fluid of strings (which is composed of Nambu strings)
  to obtain a model for a ``fluid of vortices.'' With this model, and following
  closely Soleng's proposal of a fluid of strings as the source of a
  Milgrom-type correction to the Newton dynamics, we determine quantitatively
  the modified dynamics generated by a static, spherical fluid of vortices.
\end{abstract}
\pacs{11.27.+d, 98.80.Cq, 04.20.-q \hfill gr-qc/0006048}
}

Since its development more than eighty years ago, Einstein's theory of General
Relativity has enjoyed many successes in describing gravity at very different
scales, from the orbit of Mercury to the prediction of black holes or the Big
Bang model. The theory, however, encounters serious problems when confronted to
the motion of galaxies and galaxy clusters, where it seems to imply that there
is more mass than is observed. This is the so-called ``missing mass'' problem,
to which several solutions are being investigated: first, General Relativity
may have to be modified at such distances; to conform to the observations, the
modified theory would then have to yield a gravitational acceleration
decreasing as $1/r$. Second, the Universe may be filled with dark matter, which
would act gravitationally but would not be observable. Third, non-gravitational
forces may play an important r\^ole at very large scales.  Finally, of course,
it might also prove necessary to invoke a combination of several of these
solutions.

There are good reasons, however, to believe that General Relativity is not the
definitive theory of gravitation, in that it is not a quantum theory.
Currently, string/M-theory is seen by many as the best candidate to unify the
four fundamental forces, and it is well known that in such theories Einsteinian
gravity must be replaced by more complicated scalar-tensor models.  In
low-energy M-theory, for instance, one must compactify seven of the eleven
spacetime dimensions to obtain General Relativity from eleven-dimensional
supergravity. Although it is widely believed that the extra dimensions only
play a significant r\^ole at very short distances, there exist models where
gravity becomes higher dimensional~\cite{Ruth} at very large distances as well.

A rather simple, but effective, approach to the problem is to modify the Newton
force at large distance. Milgrom~\cite{Milgrom1} proposed to write the real
gravitational acceleration $\vec{g}$ as a function of the Newtonian
acceleration $\vec{g}_{\rm N}$,
\be
  \mu \left( \frac{g}{a_0} \right) \vec{g} = \vec{g}_{\rm N},
\ee
where $a_0$ is a constant, and $\mu(x) \approx 1$ for $x \ll 1$ and $\mu(x)
\approx x$ for $x \gg 1$. This modified Newtonian dynamics was used to
explain, without the need for any dark matter, the observed gravitational
behavior of galaxies and galaxy clusters~\cite{Milgrom23,Begeman}. In
particular, it yields the a constant velocity curve at large galactic radius,
$V^4 \simeq MGa_0$, which implies that the true acceleration at large distances
is proportional to $1/r$ rather than $1/r^2$.

The main problem with this model is that it is supported only by its
phenomenological success and has no theoretical basis.

By using a model of a fluid of ordered strings, developed some time ago by
Letelier~\cite{PSL1,PSL2}, and usual General Relativity, Soleng~\cite{Sol}
found that a perfect fluid surrounding a point mass $M$ leads to a force
\be
  g = \frac{M}{r^2} + \frac1{\ell (\tilde\alpha - 2)} \left(
      \frac\ell r \right)^{1 - 2/\tilde\alpha}.
\ee
where $\ell$ is an integration constant. The parameter $\tilde\alpha$
characterizes the fluid's equation of state, and remains unspecified.

In this Letter, our main goal is to introduce a new model describing a
``perfect fluid of vortices,'' which is a relative of the well-known string
fluid~\cite{PSL2}, except that we replace the Nambu strings by (thick)
solutions of the full-field equations of the Abelian--Higgs model, namely
Nielsen--Olesen (NO) vortices~\cite{NO}. Because we choose a particle model to
describe the vortices, we are able to choose more realistically the parameter
$\tilde\alpha$ appearing in Soleng's solution, and therefore also to determine
the modified dynamics of the perfect vortex fluid surrounding a mass $M$.

Let us begin by describing the original string fluid model introduced
in~\cite{PSL2} (see also \cite{PSL1}). The starting point is the observation
that the natural generalization of the perfect fluid energy-momentum tensor to
the case of one dimension extended objects with equations of state ``tension
equals energy density'' is
\bea
  T^{\mu\nu} &=&\rho \left( u^\mu u^\nu - \chi^\nu \chi^\mu \right) 
               -p \left[g^{\mu\nu} - 
               \left( u^\mu u^\nu - \chi^\nu \chi^\mu \right)\right],
               \nonumber \\
                &=&(\rho+p)\left( u^\mu u^\nu - \chi^\nu \chi^\mu \right) -
            p  g^{\mu\nu} . \label{femt}
\eea
where $\rho$ is the string energy-density that is equal to its tension, $p>0 \;
\;(p<0)$ is a transversal pressure (tension), and $u^\mu$ and $\chi^\mu$ are
two smooth vector fields satisfying
\bml\bea
  u^\mu u^\mu &=& - \chi^\mu \chi_\mu = 1 \\
  u^\mu \chi_\mu &=& 0.
\eea\eml
At any spacetime point $u^\mu$ represents the string velocity and $\chi^\mu$
the string direction at each point of the string.

The metric for a static spherically symmetric spacetime is
\be \label{Sph}
  ds^2 = e^{2\nu(r)} dt^2 - e^{2\lambda(r)} dr^2 - r^2
         \left(d\theta^2 + \sin^2\theta d\varphi^2 \right).
\ee
The general solution of the Einstein equations for the energy-momentum
tensor~(\ref{femt}) and metric~(\ref{Sph}) was studied in~\cite{PSL2}. For the
equation of state
\be \label{eqs}
  \rho = - \, \tilde\alpha \, p,
\ee
it was found (for $\tilde\alpha \neq 2$) that
\bml\bea
  e^{2\nu} = e^{-2\lambda} &=& 1 - \frac{2M}r - \frac{\tilde\alpha}
             {\tilde\alpha - 2} \left(\frac{\ell}r\right)^{2/\tilde\alpha}, \\
  8 \pi \rho &=& \frac1{r^2} \left(\frac{\ell}r\right)^{2/\tilde\alpha}.
\eea\eml
where $\ell$ is an integration constant. This metric represents a bunch of
radially directed Nambu strings interacting transversally, in a way described
by the equation of state~(\ref{eqs}).

Our aim now is to model this transversal interaction by considering a particle
model of the strings in the fluid. More precisely, we consider an
Abelian--Higgs model with matter Lagrangian
\be
  \tilde{{\cal L}}_{\rm M}
    = \left( {\cal D}_\mu \Phi \right) \left( {\cal D}^\mu \Phi
    \right)^\dagger - \frac14 \tilde{F}_{\mu\nu} \tilde{F}^{\mu\nu} -
    \frac{\lambda} 4 \left( \Phi^\dagger \Phi
    - \eta^2 \right)^2,
\ee
where $\Phi$ is a complex Higgs field, ${\cal D}_\mu$ is a gauge-covariant
derivative defined by ${\cal D}_\mu \Phi = \nabla_\mu \Phi + ie A_\mu \Phi$ and
$\tilde{F}_{\mu\nu}$ is the strength field associated with the gauge potential
$A_\mu$, that is, $\tilde{F}_{\mu\nu} = 2 \partial_{[\mu} A_{\nu]}$.

We can simplify this theory by making the conventional Ansatz in cylindrical
coordinates $\{t, z, R, \phi\}$
\bml\bea
  \Phi   &=& \eta \, X(R) \, e^{i N \phi} \\
  A_\mu  &=& \frac1e \left( P_\mu -\nabla_a \chi \right).
\eea\eml
We can simplify this further by simultaneously rescaling our coordinates and
$P_\mu$ by the string's width $w_{\rm H} \equiv 1 / (\sqrt{\lambda} \eta)$.
This corresponds to measuring distances in string rather than Planck units, and
can be most easily done by letting $w_{\rm H} = 1$ everywhere. The only
parameter left in this theory is the Bogomol'nyi parameter
\be \label{Bogomolnyi}
  \beta \equiv \left( \frac{m_{\rm Higgs}}{m_{\rm gauge}} \right)^2 =
    \frac{\lambda}{2e^2}.
\ee

\begin{figure}[htbp]
  \centerline{\psfig{file=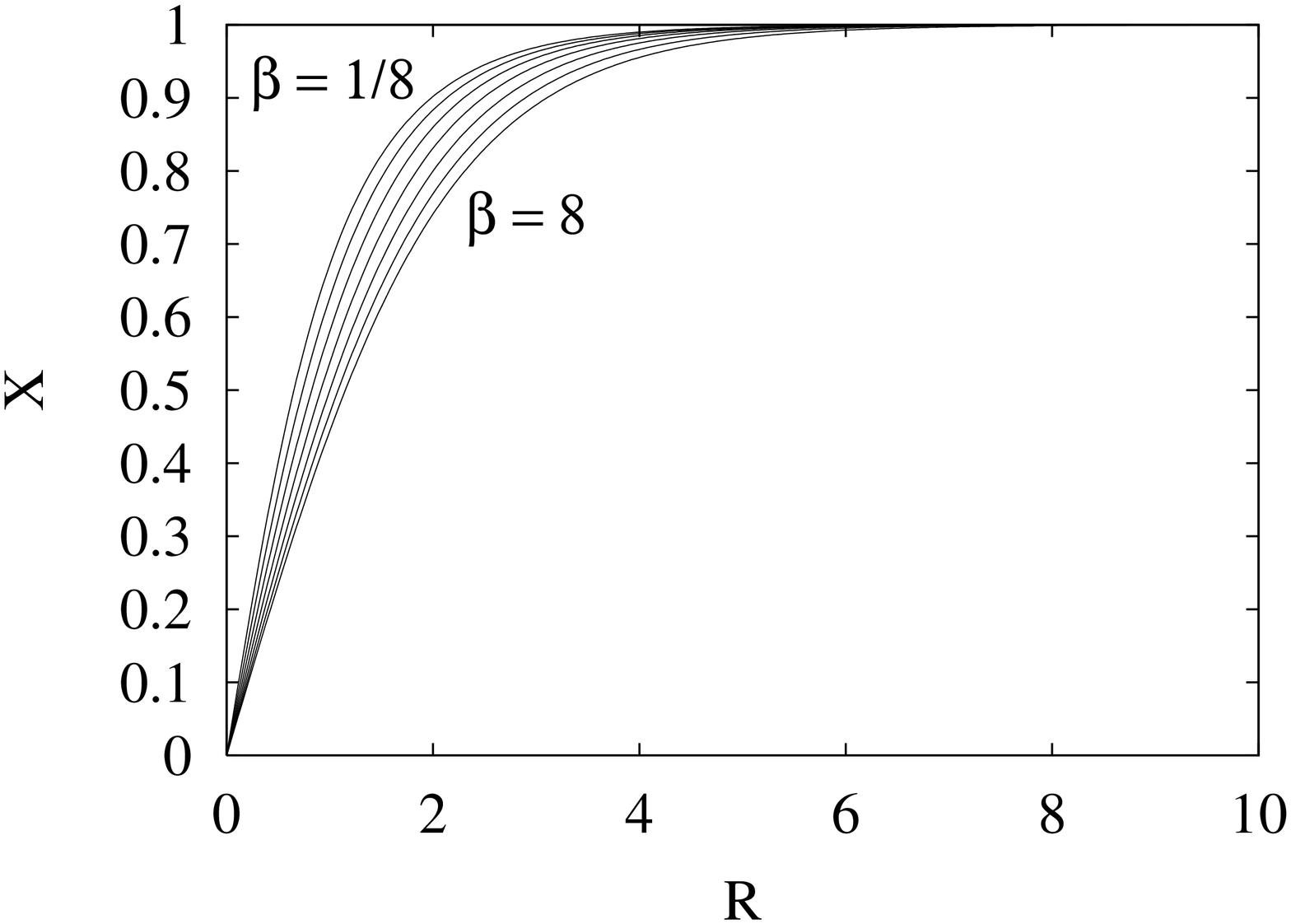,width=6.8cm,angle=270}}
  \centerline{\psfig{file=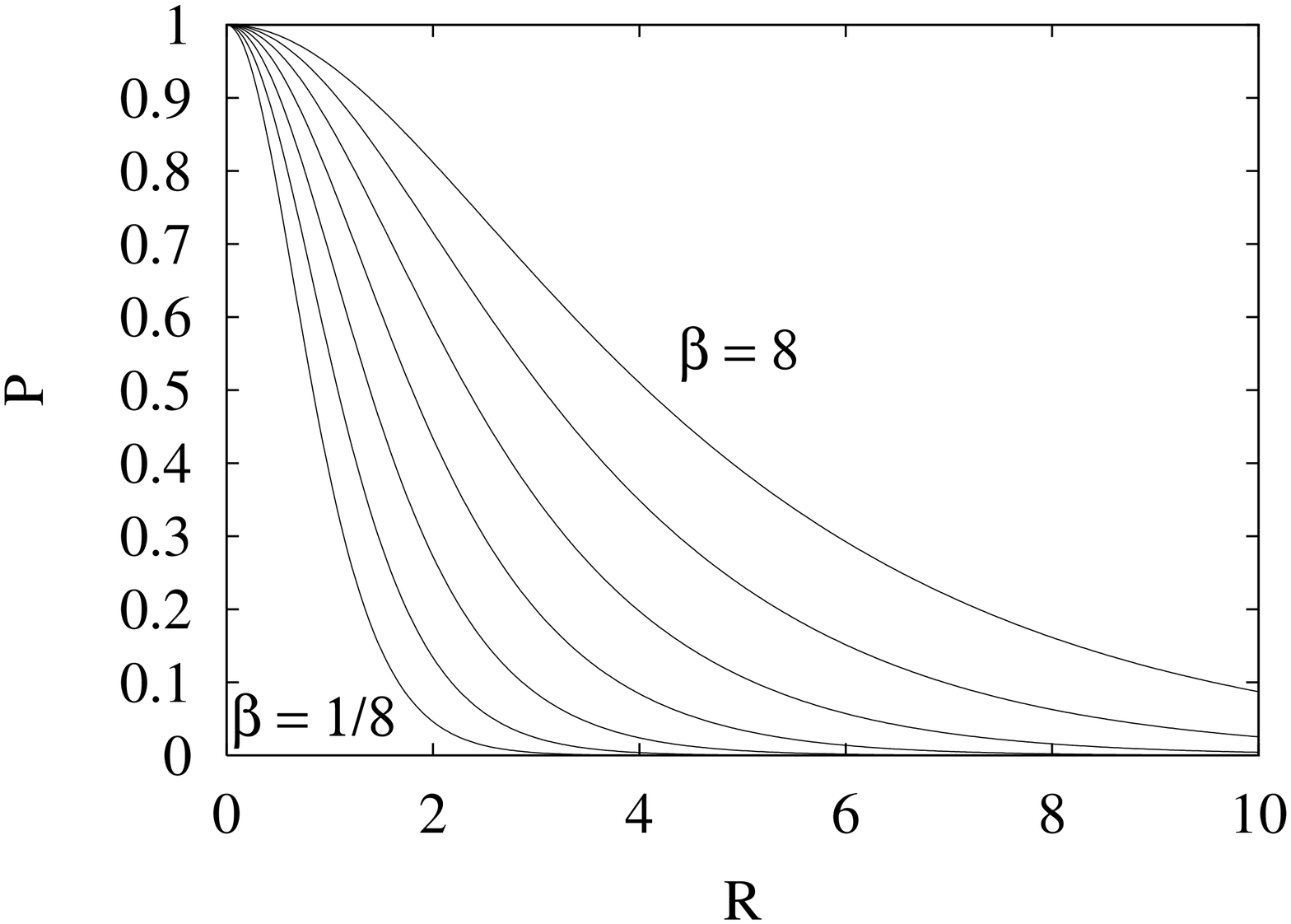,width=6.8cm,angle=270}}
  \caption{The Nielsen--Olesen solution $X(R)$ and $P(R)$ for
           a few values of the Bogomoln'yi parameter, $\beta = 1/8$, $1/4$,
           $1/2$, $1$, $2$, $4$ and $8$.}
  \label{fig:NO}
\end{figure}

The NO solution is found by assuming that $N = 1, P_\mu = P(R) \nabla_\mu
\phi$, in which case the Lagrangian and equations of motion become (after
removal of a global multiplying constant)
\be \label{eq:lag_resc}
  - {\cal L}_{\rm M} = X'{}^2 + \beta \frac{P'{}^2}{R^2} + \frac{X^2
    P^2}{R^2} + \frac14 \left( X^2 - 1 \right)^2
\ee
and
\bml \label{eq:NO} \bea
  X'' + \frac{X'}{R} - \frac{X^2 P^2}{R^2} - \frac12 X
    \left(X^2 - 1 \right) &=& 0 \\
 P'' - \frac{P'}{R} - \frac 1 \beta X^2 P &=& 0,
\eea\eml
where a prime denotes differentiation with respect to $R$. To complement these
equations, we must set the appropriate boundary conditions,
\be
  X(0) = P(\infty) = 0, \qquad \qquad X(\infty) = P(0) = 1.
\ee
With this we can solve the NO equations~(\ref{eq:NO}) numerically for various
values of the parameter $\beta$. Sample solutions are plotted on
figure~\ref{fig:NO}.

\begin{figure}[htbp]
  \centerline{\psfig{file=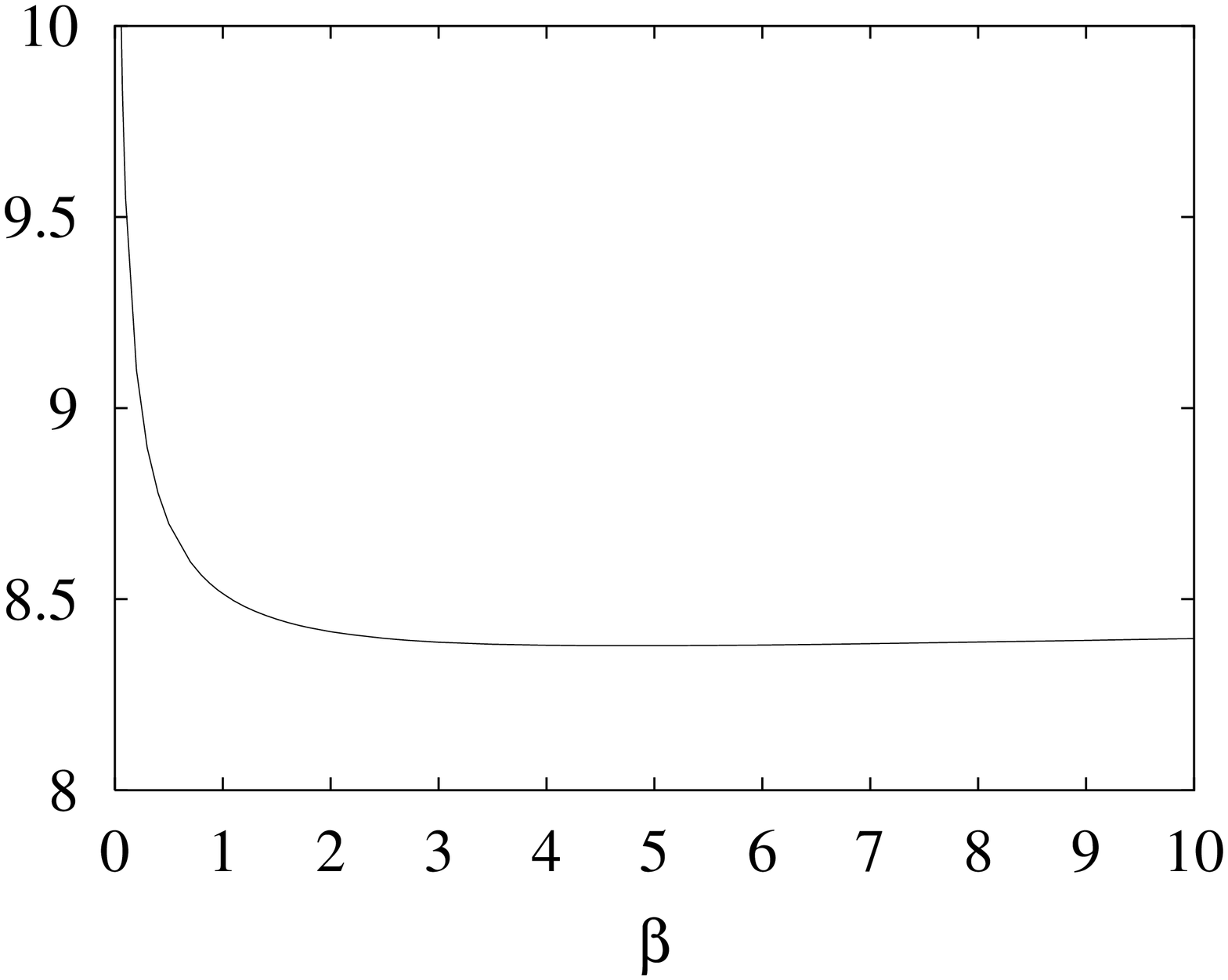,width=6.8cm,angle=270}}
  \centerline{\psfig{file=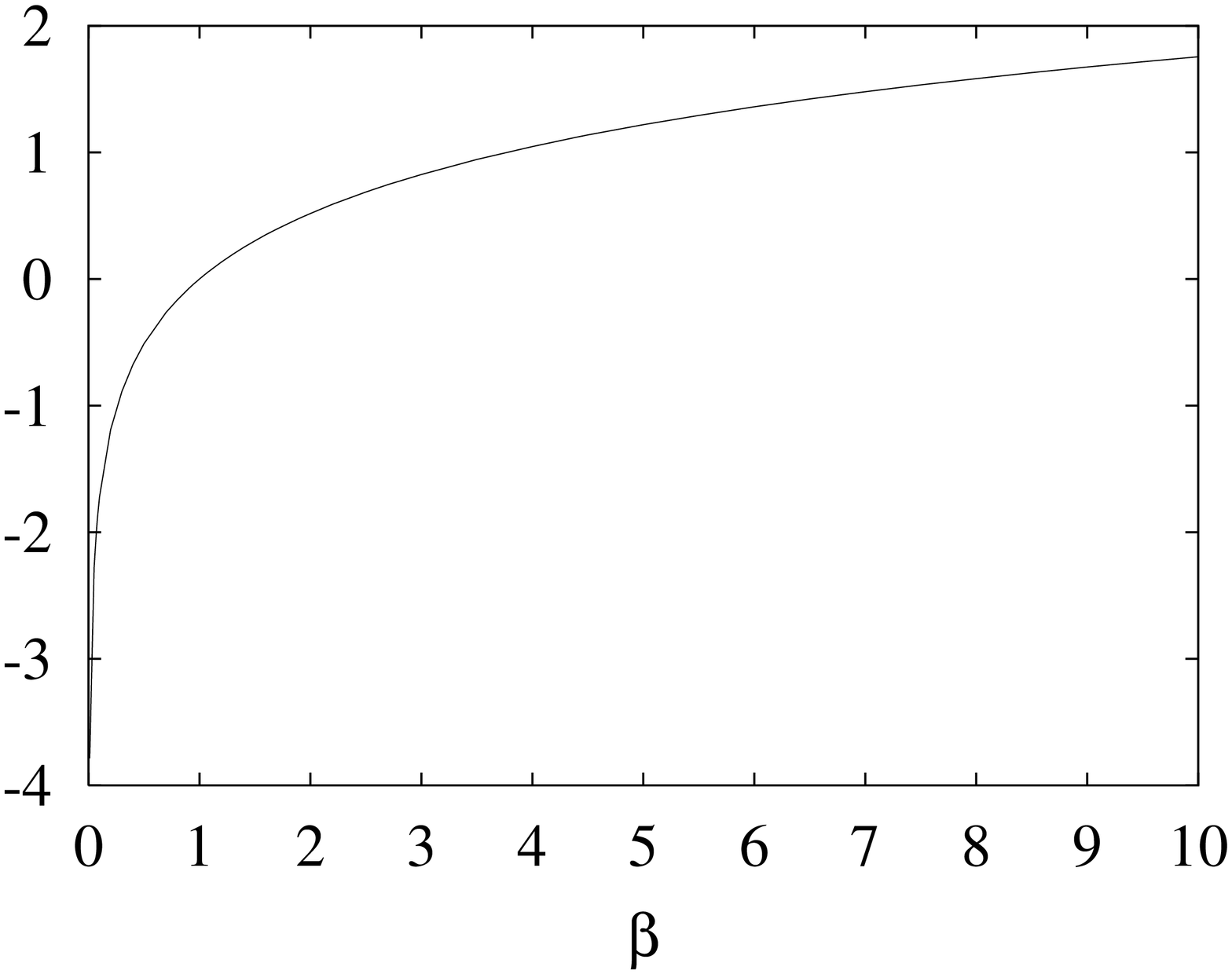,width=6.8cm,angle=270}}
  \centerline{\psfig{file=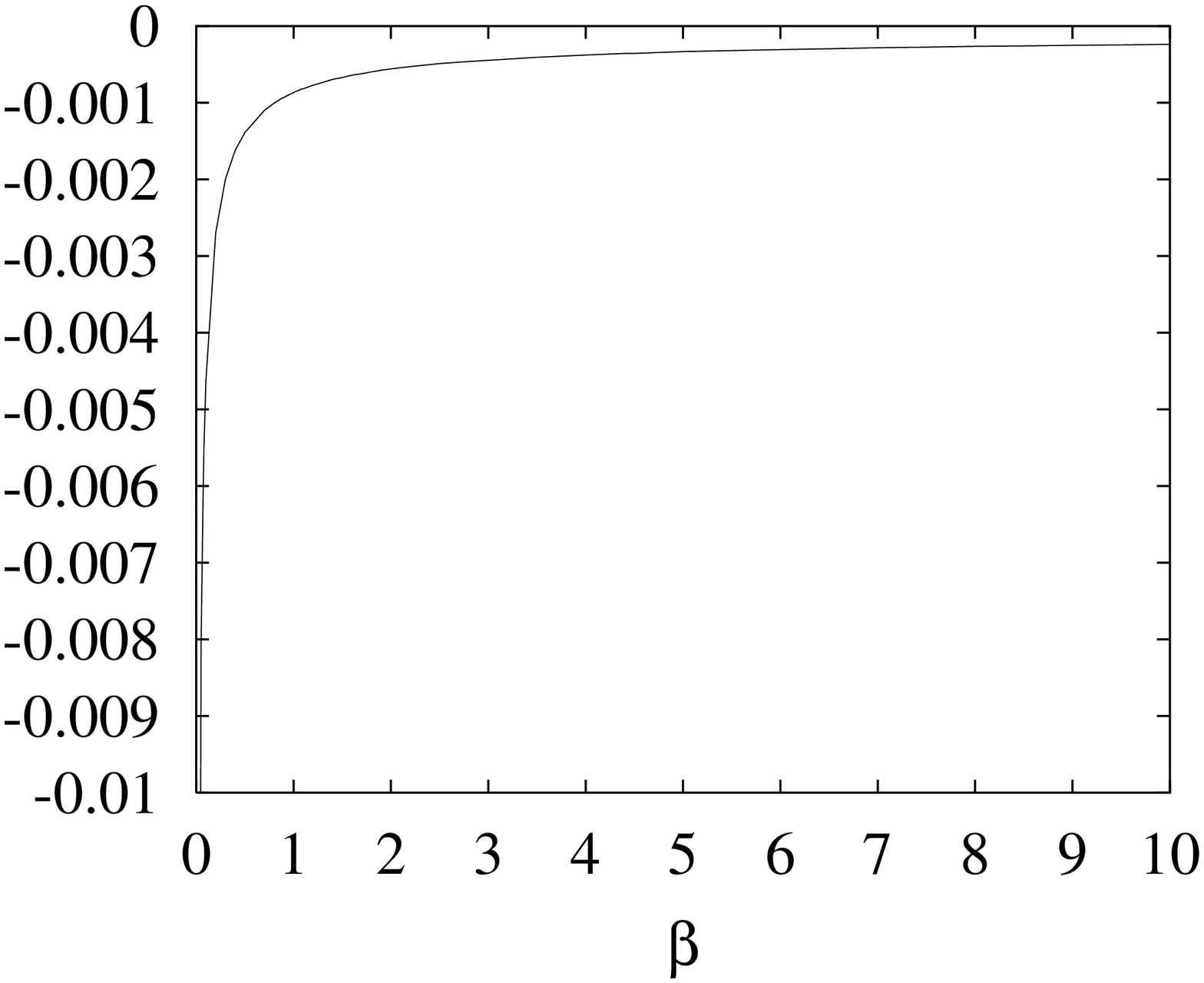,width=6.8cm,angle=270}}
  \caption{Total energy, radial pressure and azimuthal
           pressure (top to bottom) for the NO solution, as
           functions of $\beta$.}
  \label{fig:total}
\end{figure}

The energy-momentum tensor for the NO solution is given
from~(\ref{eq:lag_resc}) by
\be
  T^\mu{}_\nu = {\rm Diag} \, \left( \rho, p_z, p_R, p_\phi \right),
\ee
where
\bml\bea
  \rho   &=& \phantom{-} X'{}^2 + \beta \frac{P'{}^2}{R^2} +
             \frac{X^2P^2}{R^2} + \frac14 \left(X^2 - 1
             \right)^2 \\
  p_z    &=& -\rho = {\cal L}, \\
  p_R    &=& - X'{}^2 - \beta \frac{P'{}^2}{R^2} +
             \frac{X^2P^2}{R^2} + \frac14 \left(X^2 - 1
             \right)^2 \\
  p_\phi &=& \phantom{-} X'{}^2 - \beta \frac{P'{}^2}{R^2} -
             \frac{X^2P^2}{R^2} + \frac14 \left(X^2 - 1
             \right)^2.
\eea\eml
Note that for $\beta = 1$ the equations of motion reduce to the Bogomol'nyi
equations,
\be
  X' = \frac{XP}{R}, \qquad \qquad P' = \frac{R}2 \left(X^2 -
1 \right),
\ee
which yields $p_R = p_\phi = 0$. For the purpose of finding the vortex's
equation of state, we must integrate the energy and the pressures; these are
shown in figure~\ref{fig:total}.

\begin{figure}[htbp]
  \centerline{\psfig{file=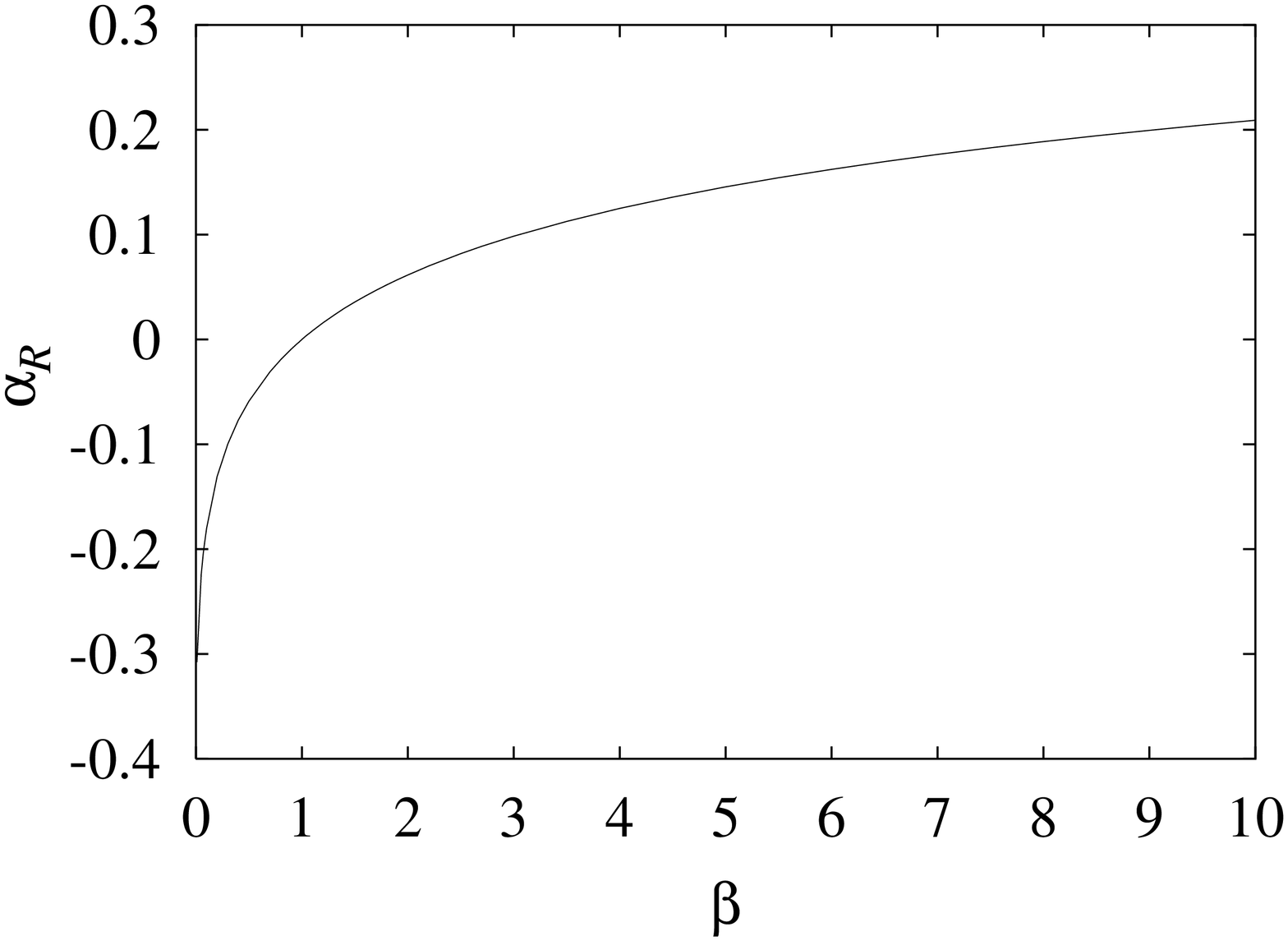,width=6.8cm,angle=270}}
  \centerline{\psfig{file=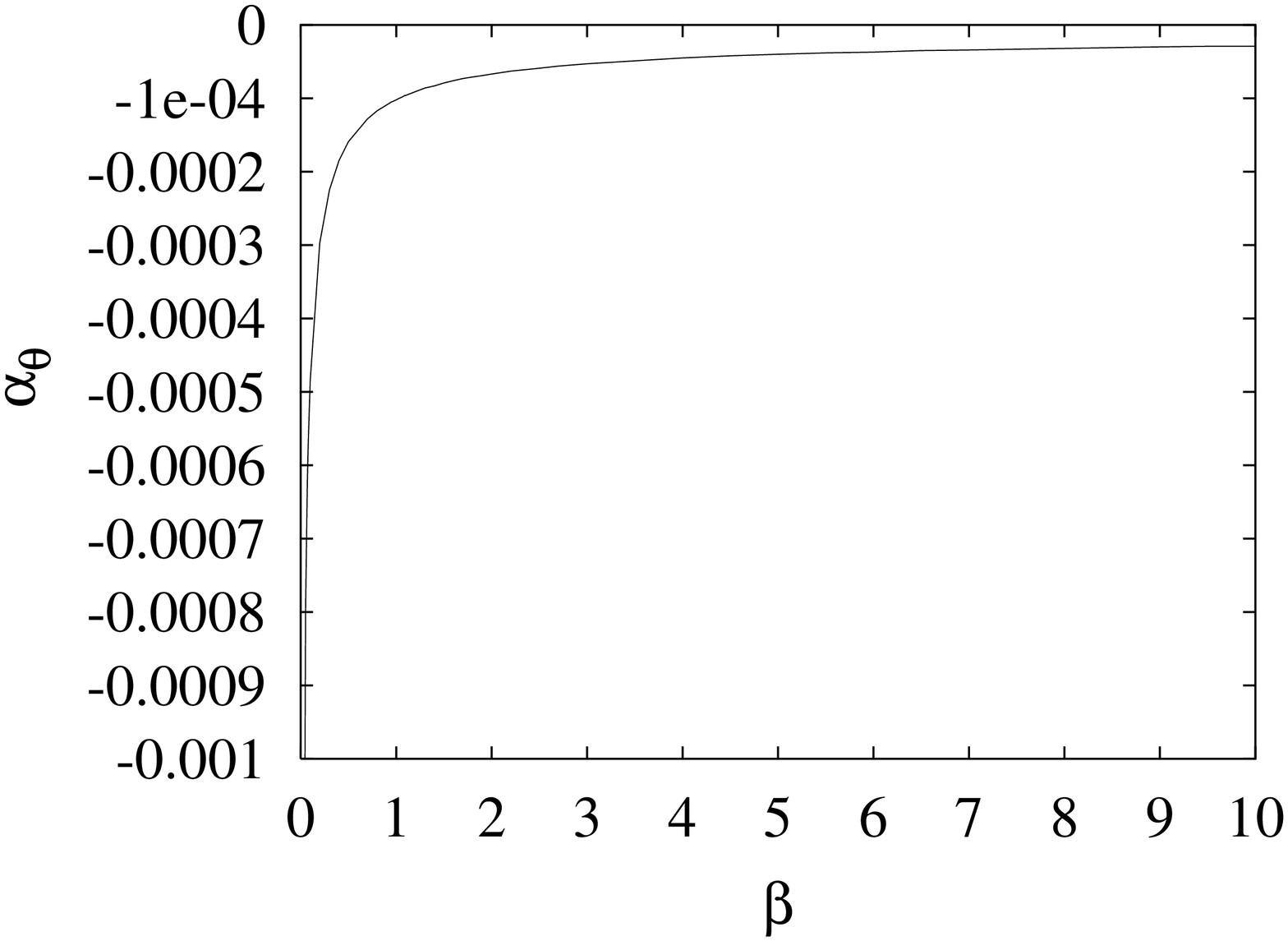,width=6.8cm,angle=270}}
  \caption{The coefficients $\alpha_R$ and $\alpha_\theta$ of the equations of
           state for the NO vortex, as functions of $\beta$.}
  \label{fig:state}
\end{figure}

Finally, we can write the equations of state of the vortex as
\bml\bea
  p_R        &=& \alpha_R(\beta)  \rho \\
  p_{\theta} &=& \alpha_\theta(\beta) \rho.
\eea\eml
The functions $\alpha_R(\beta)$ and $\alpha_\theta(\beta)$ are given
on figure~\ref{fig:state}.

We are now in a position to determine realistic values for $\alpha_R$ and
$\alpha_\theta$. Because the Abelian--Higgs theory is a toy model, there is no
experimental determination of the coefficients $\lambda$ and $e$ appearing in
the definition of $\beta$, but we can compute it from the masses of the Higgs
and the gauge boson. We assume here that
\be
  77 < m_{\rm H} {\rm (GeV)} < 300
\ee
and that the most probable value is $m_{\rm H} \approx 170$~GeV~\cite{Masses}.
For the gauge boson, we assume a $W^\pm$ or a $Z^0$, with
\be
  m_{\rm W} = 80 {\rm GeV}, \qquad \qquad
  m_{\rm Z} = 82 {\rm GeV}.
\ee
This yields
\bml\bea
   0.88                \lesssim &~\beta~&         \lesssim 14.0  \\
  -0.011               \lesssim &~\alpha_R~&      \lesssim 0.239 \\
  -1.103 \times 10^{-4}\lesssim &~\alpha_\theta~& \lesssim -2.51 \times
                                                           10^{-5},
\eea\eml
the most probable value for $m_{\rm H}$ implying $\beta = 4.4$, $\alpha_R =
0.134$ and $\alpha_\theta = -4.27\times 10^{-5}$.

We can now apply what we found to a spherical domain with metric~(\ref{Sph})
containing a fluid of vortices, by writing the equation of state for this fluid
as
\be \label{eqsvortex}
  p = \frac12( \alpha_R + \alpha_\theta ) \rho = \kappa \rho,
\ee
where $-5.51\times 10^{-3} \lesssim \kappa \lesssim 0.119$ and most probable
value $\kappa = 0.067$. Insertion of this into Soleng's result~\cite{Sol}, via
$\kappa = -1/\tilde\alpha$, gives
\bml\bea
  e^{2\nu} &=& e^{-2\lambda} = 1 - \frac{2M}r + \frac1{2\kappa+1} \left(
               \frac{r}{r_0} \right) ^{2\kappa} \\
  \rho     &=& -\frac1{r^2} \left( \frac{r}{r_0} \right) ^{2\kappa},
\eea\eml
which yields, in the Newtonian limit, an effective acceleration of magnitude
\be \label{realg}
  g = \frac{M}{r^2} + \frac{K}{r^{1-2\kappa}},
\ee
where $K$ is a constant. Numerically, we obtain that the correction decays like
$r^{-1.011}$ (for $\beta = 0.88$), $r^{-0.866}$ (for $\beta = 4.4$) and
$r^{-0.762}$ (for $\beta = 14$). It is exactly Milgrom ($\propto 1/r$) if
$\beta = 1$.

\begin{figure}[htbp]
  \centerline{\psfig{file=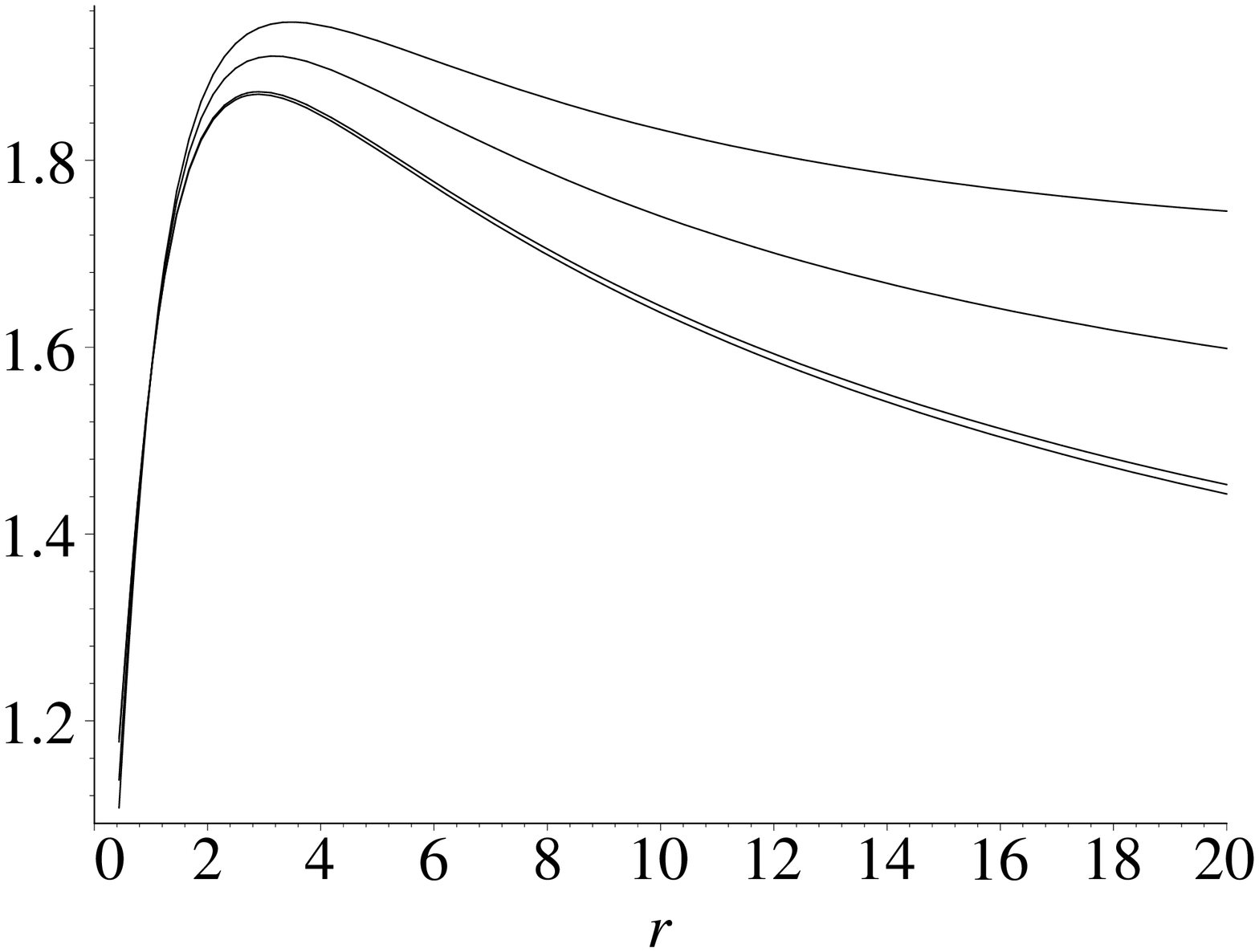,width=6.8cm,angle=270}}
  \centerline{\psfig{file=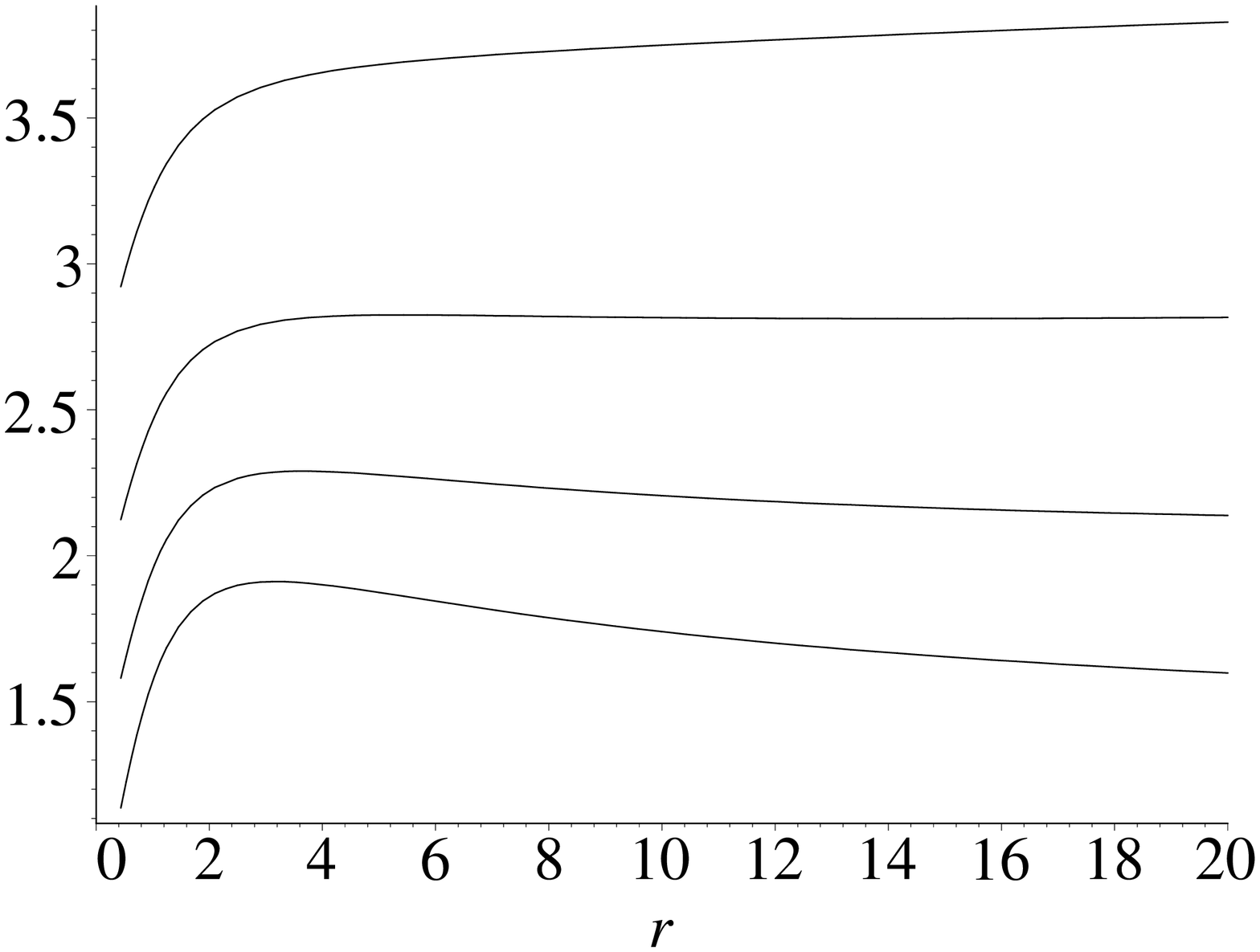,width=6.8cm,angle=270}}
  \caption{Sample rotation curves obtained from the acceleration
           (\protect\ref{realg}). On the top figure, we plot the curves for
           (bottom to top) $\beta = 0.88$, 1, 4.4 and 14. On the bottom figure,
           we plot the curves for $\beta = 4.4$ and (bottom to top) $K = 0.1$,
           0.25, 0.5 and 1.}
  \label{fig:rot}
\end{figure}

Lastly, we compute the rotation curves for our modified dynamics, assuming a
modified Hubble profile~\cite{GD} to find $M(R)$. In figure~\ref{fig:rot} we
plot the rotation curves for $K = 0.1$ and several values of $\beta$, as well
as the curves for $\beta = 4.4$ for several values of $K$. The figure shows
that, within the current limits on the Higgs mass, the modified dynamics allows
for a broad family of curves, depending on the parameter $K$. Milgrom-type
corrections are also allowed, although an exactly Milgrom $1/r$ dependence
seems rather unlikely.

Although more tests would clearly be needed, the existence of a model such as
the one presented in this Letter shows that one could imagine a situation,
compatible with observations of the dynamics of galaxies and galaxy clusters,
where dark matter would not be needed at all (or, taking a different point of
view, would consist in our case of a fluid of vortices). This would have the
advantage, over other dark matter models, that it is based on a particle theory
(albeit a toy model here), giving it a stronger theoretical base.

\section*{Acknowledgements}

F.B. is grateful to Stephen Burby and Ruth Gregory for useful discussions. The
authors wish to thank FAPESP and CNPq for financial support.

%%
%% Bibliography
%%

\def\cmp#1 #2 #3.{{\it Commun.\ Math.\ Phys.\ \bf#1} #2 (#3).}
\def\cqg#1 #2 #3.{{\it Class.\ Quantum Grav.\ \bf#1} #2 (#3).}
\def\mpla#1 #2 #3.{{\it Mod.\ Phys.\ Lett.\ \rm A\bf#1} #2 (#3).}
\def\ncim#1 #2 #3.{{\it Nuovo Cim.\ \bf#1\/} #2 (#3).}
\def\npb#1 #2 #3.{{\it Nucl.\ Phys.\ \rm B\bf#1} #2 (#3).}
\def\plb#1 #2 #3.{{\it Phys.\ Lett.\ \bf#1\/}B #2 (#3).}
\def\pr#1 #2 #3.{{\it Phys.\ Rev.\ \bf#1} #2 (#3).}
\def\prd#1 #2 #3.{{\it Phys.\ Rev.\ \rm D\bf#1} #2 (#3).}
\def\prl#1 #2 #3.{{\it Phys.\ Rev.\ Lett.\ \bf#1} #2 (#3).}
\def\grg#1 #2 #3.{{\it Gen. Rel. Grav.\ \bf#1} #2 (#3).}
\def\apj#1 #2 #3.{{\it Ap. J.\ \bf#1} #2 (#3).}
\def\mnras#1 #2 #3.{{\it Mon. Not. R. astr. Soc.\ \bf#1} #2 (#3).}


\begin{references}
  \bibitem[*]{bonj}      E-mail: \texttt{bonjour@ime.unicamp.br}.
  \bibitem[\dagger]{psl} E-mail: \texttt{letelier@ime.unicamp.br}.
  \bibitem{Ruth}         R. Gregory, Valery A. Rubakov and Sergei
                         M. Sibir\-yakov, [hep-th/0002072].
  \bibitem{Milgrom1}     M. Milgrom, \apj 270 365 1983.
  \bibitem{Milgrom23}    M. Milgrom, \apj 270 371 1983. \apj 270 384 1983.
  \bibitem{Begeman}      K.G. Begeman, A.H. Broeils and R.H. Sanders, \mnras
                         249 523 1991.
  \bibitem{PSL1}         P.S. Letelier, \prd 20 1294 1978.
  \bibitem{PSL2}         P.S. Letelier, \ncim 63 519 1981.
  \bibitem{Sol}          H.H. Soleng, \grg 27 367 1995.
  \bibitem{NO}           H.H. Nielsen and P. Olesen, \npb 61 45 1973.
  \bibitem{Masses}       G. D'Agostini and G. Degrassi [hep-ph/9902226].
                         P. Gambino [hep-ph/9812332]. J. Ellis
                         [hep-ph/9507424].
                         C. Caso {\it et. al} (Particle Data Group) {{\it Eur.
                         Phys. J.} {\bf C3} 1 (1998).}
  \bibitem{GD}           James Binney and Scott Tremaine, ``Galactic
                         Dynamics,'' Princeton University Press (NJ: Princeton)
                         1987.
\end{references}
\end{document}